\documentclass[prx,aps,showpacs,twocolumn,longbibliography,superscriptaddress]{revtex4-2}
\usepackage{xcolor}
\usepackage{color} %used for font color 
\usepackage{amssymb} %maths            
\usepackage{amsmath} %maths     
\usepackage{braket}
\usepackage{graphicx}
\usepackage{epstopdf}
\usepackage[english]{babel}
\usepackage{amsthm}
\usepackage{amsfonts}
\usepackage{hyperref}
\newcommand{\be}{\begin{equation}}
\newcommand{\ee}{\end{equation}}
\newcommand{\bea}{\begin{eqnarray}}
\newcommand{\eea}{\end{eqnarray}}
\newcommand{\ba}{\begin{array}}
\newcommand{\ea}{\end{array}}

\begin{document}

\title{High Harmonic Tracking of Ultrafast Electron Dynamics across the Mott to Charge Density Wave Phase Transition}

\author{Marlena Dziurawiec}
\affiliation{Institute of Theoretical Physics, Wroc{\l}aw University of Science and Technology, 50-370 Wroc{\l}aw, Poland}

\author{Jessica O. de Almeida}
\affiliation{ICFO - Institut de Ci\`encies Fot\`oniques, The Barcelona Institute of Science and Technology, 08860 Castelldefels (Barcelona), Spain.}

\author{Mohit Lal Bera}
\affiliation{Departamento de F\'{i}sica Te\'{o}rica and IFIC,  Universidad de Valencia-CSIC, 46100 Burjassot (Valencia), Spain}
\affiliation{ICFO - Institut de Ci\`encies Fot\`oniques, The Barcelona Institute of Science and Technology, 08860 Castelldefels (Barcelona), Spain.}

\author{Marcin Płodzień}
\affiliation{ICFO - Institut de Ci\`encies Fot\`oniques, The Barcelona Institute of Science and Technology, 08860 Castelldefels (Barcelona), Spain.}

\author{Maciej Lewenstein}
\affiliation{ICFO - Institut de Ci\`encies Fot\`oniques, The Barcelona Institute of Science and Technology, 08860 Castelldefels (Barcelona), Spain.}
\affiliation{ICREA, Pg. Lluis Companys 23, ES-08010 Barcelona, Spain.}

\author{Tobias Grass}
\affiliation{DIPC - Donostia International Physics Center, Paseo Manuel de Lardiz{\'a}bal 4, 20018 San Sebasti{\'a}n, Spain}
\affiliation{IKERBASQUE, Basque Foundation for Science, Plaza Euskadi 5, 48009 Bilbao, Spain}

\author{Ravindra W. Chhajlany}
\affiliation{ISQI - Institute of Spintronics and Quantum Information, Adam Mickiewicz University, 61-712 Pozna\'n, Poland.}

\author{Maciej M. Ma\'ska}
\affiliation{Institute of Theoretical Physics, Wroc{\l}aw University of Science and Technology, 50-370 Wroc{\l}aw, Poland}

\author{Utso Bhattacharya}
\affiliation{Institute for Theoretical Physics, ETH Zurich, 8093 Zurich, Switzerland}
\affiliation{ICFO - Institut de Ci\`encies Fot\`oniques, The Barcelona Institute of Science and Technology, 08860 Castelldefels (Barcelona), Spain.}

\begin{abstract}
Different  insulator phases compete with each other in strongly correlated materials with simultaneous local and non-local interactions. It is known that the homogeneous Mott insulator converts into a charge density wave (CDW) phase when the non-local interactions are increased, but there is ongoing debate on whether and in which parameter regimes this transition is of first order, or of second order with an intermediate bond-order wave phase.   Here we show that strong-field optics applied to an extended Fermi-Hubbard system can serve as a powerful tool to reveal the nature of the quantum phase transition. Specifically, we show that in the strongly interacting regime characteristic excitations such as excitons, biexcitons, excitonic strings, and charge droplets can be tracked by the non-linear optical response to an ultrafast and intense laser pulse. 
Subcycle analysis of high harmonic spectra unravels the ultrafast dynamics of these increasingly complex objects, which partially escape the scrutiny of linear optics. Their appearance in the high harmonic spectrum provides striking evidence of a first-order transition into the CDW phase, and makes a strong case for using strong-field optics as a powerful tool to reveal the nature of quasiparticles in strongly correlated matter, and to track the electron dynamics during a first-order quantum phase transition.  
\end{abstract}

\maketitle

\section{Introduction} 

Advances in laser technology, facilitating the creation of high-intensity and ultrashort optical pulses \cite{RevModPhys.91.030501, RevModPhys.91.030502}, have ushered in a new era of exploring matter through non-linear optical responses. High-harmonic generation (HHG) has emerged as a focal point in this quest. This process leads to the ultrafast emission of radiation with frequencies as integer multiples of the initial driving frequency. While HHG in atomic gases has long been recognized as a pivotal source of extreme ultraviolet attosecond pulses \cite{nobel2023,hentschel_attosecond_2001, paul_observation_2001, ferray_multiple_1988, brabec2000}, driving the foundational principles of attosecond physics \cite{Amini_2019, Krausz_Ivanov_2009, Krause_1992, corkum_plasma_1993}, the past decade has witnessed experimental breakthroughs that extend HHG investigations to bulk solids. This expansion into condensed matter systems has ignited a surge of theoretical and experimental inquiries, offering a brand new tool for unraveling intricate electronic properties within these materials \cite{Vampa2015, Vampa_2017}. In drawing parallels between atomic and solid-state HHG, the theoretical foundation of both processes is the semiclassical single-active-electron three-step model \cite{corkum_plasma_1993,lewenstein_theory_1994, Kulander1993}. Originating from the atomic perspective, it appropriately captures the intricate sequence of events, describing how electrons tunnel, accelerate, and recombine, leading to high-harmonic emission. Adapted to solids, the three-step model accounts for the crystal's band structure, in which the electrons traverse the valence and conduction bands, creating and recombining with holes, producing a harmonic spectrum with distinctive features which encompass two distinct processes — intraband motion associated with lower-order harmonics and interband recombination dominating in the high-order harmonics regime. This unified framework bridges the gap between atomic and solid-state HHG. Due to its high temporal resolution, high harmonic spectroscopy provides the ability to probe ultrafast electronic motion in solid state systems \cite{Ghimire2011,Schubert2014,Hohenleutner2015,Vampa2015,Ndabashimiye2016,Liu2017}. This has ignited substantial theoretical and experimental investigations, recently including quantum electrodynamical aspects \cite{Bhattacharya2023}. These advancements in bulk material experimentation have spurred the development of methodologies to discern dynamical properties of Bloch electrons, including the detection of electronic band structures \cite{Luu2015,vampa_all-optical_2015,li2020}, Berry curvature \cite{Luu2018}, transition dipole moment \cite{Uchida2021} and topology \cite{Sato2019,Silva2019,Chaicon2020,Jurss2019,Jurss2020,baldelli_detecting_2022,Bera2023,Neufeld2023}. Despite the extensive study of HHG in semiconductors and semimetals based on a single-particle band picture, recent attention has shifted toward unraveling the intricacies of many-body effects on HHG in strongly correlated systems, including recent experimental investigation of high-$T_c$ cuprate superconductors \cite{Alcala2022}.

\begin{figure}[t]
    \centering
    \includegraphics[width=1\columnwidth]{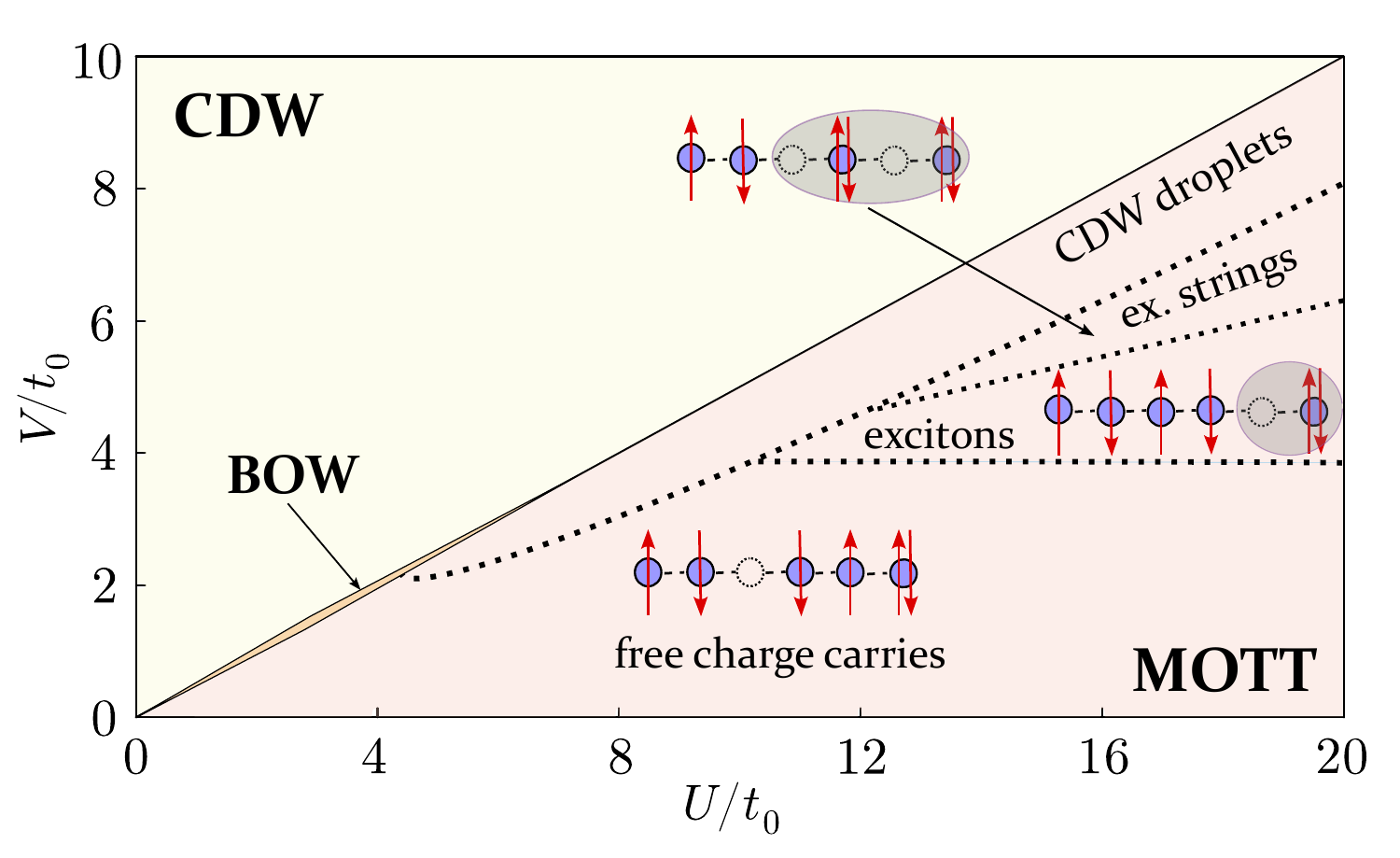}
    \caption{Schematic phase diagram of the extended Fermi-Hubbard model, exhibiting Mott insulating phase, charge-density wave (CDW) phase, and bond-order wave (BOW) phase \cite{zhang_dimerization_2004}. When the extended interactions $V$ are increased in the strongly interacting regime (i.e. at large $U/t_0$), more and more complex optical excitations are formed, and can be interpreted as precursors of the first-order Mott-CDW transition \cite{jeckelmann_optical_2003}.
    %\textcolor{blue}{[Add some Refs.?]}
    }
    \label{fig:excitation_diagram}
\end{figure}

In the realm of strong electronic correlations, the conventional single-particle framework mentioned earlier proves inadequate. A vital minimal model is the one-dimensional (1D) extended Fermi Hubbard model (EFHM), incorporating a competition between hopping $t_0$ and both on-site repulsion $U$ and nearest-neighbor repulsion $V$:
\begin{align}
\label{EFHM}
        \hat{H} &= - t_0 \sum_{j,\sigma} \left( \hat{c}^{\dagger}_{j\sigma} \hat{c}_{j+1\sigma} + {\rm H.c.} \right) \nonumber \\ & + \sum_j \left( U \hat{n}_{j\uparrow} \hat{n}_{j\downarrow}  + V \hat{n}_j \hat{n}_{j+1} \right),
\end{align}
with $\hat c_{j\sigma},\:\hat c_{j\sigma}^\dagger,\:\hat n_{j\sigma}\ (\hat{n}_j=\hat{n}_{j\uparrow}+\hat{n}_{j\downarrow})$ being the fermionic creation/annihilation/number operators.
This model effectively captures the diverse physical properties of materials recognized as prototypical one-dimensional Mott insulators, including conjugated polymers (e.g., $\mathrm{ET-F_2TCNQ}$ \cite{wall_quantum_2011, okamoto_photoinduced_2007}), chain cuprates (e.g., $\mathrm{Sr_2CuO_3}$ \cite{kishida_gigantic_2000, silva_high-harmonic_2018}), or Ni halides. These materials boast significant optical non-linearity, and exhibit intricate phase diagrams which arise out of strong inter-electron interaction effects, indicating their promising potential for many applications
\cite{kiess_conjugated_1992, ogasawara_ultrafast_2000,kishida_gigantic_2000,Iwai2003}.  
The EFHM proves to be effective in portraying an interaction-induced metal-insulator phase transition, giving rise to the Mott insulator (MI) with quasi-long-range spin density wave antiferromagnetic order or charge density wave (CDW) insulating phases when $V$ dominates. The CDW phase spontaneously breaks the discrete translational symmetry of the lattice. Despite its apparent simplicity, numerical simulations of this model pose significant challenges. The phase diagram of this model at half-filling has been obtained by the density matrix renormalization group (DMRG) method \cite{zhang_dimerization_2004}. In addition to the MI and CDW, it also contains a narrow intermediate region with bond order wave (BOW) phase. However, in the strong interaction limit, a large charge gap prevents the BOW order, and when $V$ exceeds $U/2$, the system enters a CDW phase through a first-order phase transition. Deep in the Mott phase, the primary charge-conserving excitations involve generating a doubly occupied site (doublon), incurring a local energy of approximately $\sim U$, and an empty site (holon), as an unbound pair of elementary charge carriers moving independently through the lattice.
As the nearest-neighbor interaction parameter $V$ is increased, the low-energy optical excitations involve collective manifestations of these doublon-holon pairs. For $V > 2t_0$, doublon-holon pairs amalgamate to form various excitonic states. Consequently, within the Mott phase of the EFHM, the optical response differs significantly in the presence of these excitons compared to the case with $V=0$, as the motion of doublons and holons is severely confined by the interactions between these pairs \cite{jeckelmann_optical_2003}, see also Fig.~\ref{fig:excitation_diagram} for a schematic diagram. By increasing the nearest-neighbor interactions, doublon-hole pairs first bind to excitons, reducing the cost associated with larger $V$. Such an exciton can also be viewed as the elementary cell of a CDW insulator, and the first-order transition into the CDW implies the growth of collective excitations consisting of multiple doublon-hole pairs, known as excitonic strings, with well-defined sizes, such as biexcitons and triexcitons. Finally, as the system approaches the critical point ($0<U - 2V \lesssim t_0$), the low-energy excitations evolve into a superposition of excitonic strings of various sizes, termed CDW droplets. These droplets are, in fact, precursors of the CDW order observed in the ground state when $U<2V$.

Theoretically, signatures of these different excitations should be present in the linear optical response, but the smallness of the matrix elements between droplet and ground state makes it extremely hard to detect them via the linear response. In the present paper, we build up on previous work which has established high harmonic generation (HHG) as a tool to probe Hubbard-like system. Specifically, these investigations extend to both one-dimensional models \cite{silva_high-harmonic_2018, murakami_high-harmonic_2021} and to higher dimensions \cite{murakami_high-harmonic_2018, orthodoxou_high_2021}. Additionally, theoretical and experimental exploration encompasses other models of strongly correlated systems and diverse modifications of the Hubbard model \cite{alshafey_ultrafast_2022}. This also includes scrutiny of the extended Hubbard model near its critical point \cite{shao_high-harmonic_2022} and within the realm of intersite interactions, where the optical response reveals the presence of excitons, that is, bound states of doubly-occupied and empty sites \cite{udono_excitonic_2022}.
Exploring one-dimensional Mott insulators (MIs) through optical experiments, especially those that exhibit pronounced third-harmonic responses, has already brought to light the crucial role played by excitonic effects \cite{Ono2004,kishida_gigantic_2000,ogasawara_ultrafast_2000,Ono2005}. While the anticipation is that the exciton within the Mott insulator contributes to high harmonic generation (HHG) \cite{Udono2022}, it is imperative to recall that, unlike a simple Mott insulator with weak $V$, multi-exciton effects become truly dominant in materials with strong $V$. 
This is the goal of the present study which delves into the exploration of excitonic effects on HHG within the Mott phase and unravels the intricacies of optical excitations characterizing the transition from the Mott insulator phase to a charge density wave (CDW) phase. Our investigation encompasses the following key aspects: (a) elucidating the varying roles played by different excitonic effects as $V$ increases, shaping the HHG response of the system, (b) deciphering how these excitonic effects orchestrate a first-order phase transition from a Mott insulator to a CDW phase, (c) discerning the analogous role played by spin droplet excitations, akin to excitons, in determining the HHG spectrum in the CDW phase, and (d) utilizing subcycle analysis to probe the structures of excitations responsible for generating peaks in the emitted HHG signal. This comprehensive exploration enhances our understanding of the dynamic interplay between excitonic effects and high harmonic generation, shedding light on the rich phase transitions within an experimentally relevant 1D strongly correlated system.

\section{Results}
\label{s:static_HHG}
\subsection{High harmonic spectra}

\begin{figure*}[ht]
    \centering
    \includegraphics[width=2\columnwidth]{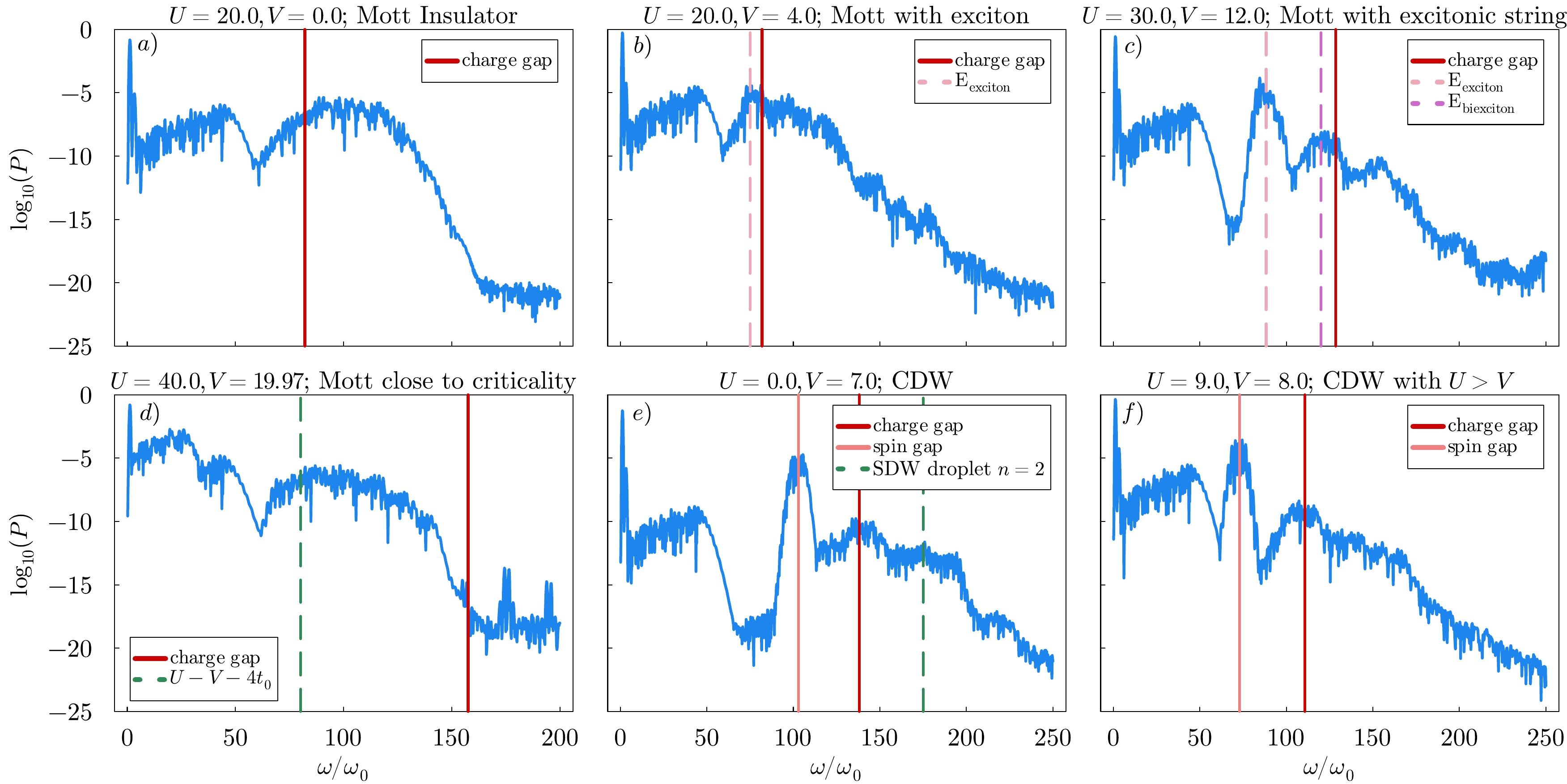}
    \caption{HHG spectra calculated for laser parameters: $A_0=4$, $\omega_0=0.2$, $n_c = 5$, system size $N=12$, for Hamiltonian \ref{EFHM} parameters corresponding to: a) Mott phase with free charge carriers, b) Mott with an exciton, c) Mott with excitonic strings, d) Mott with CDW droplets, e) CDW phase with zero-$U$, f) realistic ($U>V$) CDW phase. Vertical lines correspond to the characteristic energies of the system. The additional below-gap peak, present in all of the spectra at frequency $\simeq 50 \omega_0$ is an effect of a finite size of the system and disappears in the thermodynamic limit; see Supplementary Materials for details. 
    %{\color{red} 
    %[Comment: Maybe consider removing the $3V-U$ and $4v-U$ lines in (e,f), because they are (good) approximations to spin and charge gap? Maybe we could call $5V-2U$ "SDW droplet $n=2$?. Why are some lines dashed and others solid, why are no dashed lines in the legends? Can we remove the $A_0$ and $\omega_0$ information from the panel headlines (It's enough to have it in the caption). Instead, we could include "MI" or "CDW" or "Criticality" in the headlines, to ease the reading of the figure?
    %]
    %}\textcolor{blue}{[Should we not mention the system size here or somewhere in the main text, I guess $L=12$?]}
    }
    \label{fig:HHG}
\end{figure*}

Concentrating on the undoped and unpolarized EFHM, we have calculated high harmonic spectra across various regimes. The results are shown in Fig.~\ref{fig:HHG}. 

\paragraph{MI phase at $V=0$.}
We begin by examining the scenario characterized deep in the Mott phase, with strong on-site copuling ($U/t_0=20$) and negligible non-local interactions ($V/t_0\leq2$). The Mott insulating ground state reveals quasi-long-range antiferromagnetic order, where electron spins exhibit a preference for anti-parallel alignment at neighboring sites \cite{gebhard_mott_1997}. The elementary charge excitations are independently delocalized doublon-holon pairs, set apart from the half-filled $N$-electron ground state by the charge gap $\equiv E_0(N+1)+E_0(N-1)-2E_0(N)$, where $E_0(N)$ is the ground state energy of a system filled with $N$ particles. The creation of an excitation on top of the uniformly filled ground state, with one particle per site, incurs an energy cost on the order of $\sim U$. This insulating behavior is clearly represented in the HHG spectrum, as depicted in Fig.~\ref{fig:HHG}a). Notably, a dip at the charge gap (orange line) is discernible up to a harmonic order of $\sim (U-O(t^2_0))/\omega_0$, succeeded by a plateau. The plateau arises from the energy continuum formed by a single delocalized doublon-holon pair, which exhibits metallic behavior. The single plateau above the Mott gap concludes at the cutoff, determined by the maximal energy of a single doublon-hole pair. This observation aligns with the spectrum observed in the preceding work \cite{silva_high-harmonic_2018}.

\paragraph{Effect of finite $V$ on MI phase.}
Next, we delve into the impact of a non-zero $V<U/3$ (specifically, $U/t_0 = 20$, $V/t_0= 4$) where a particularly intriguing set of behavior unfolds. As illustrated in Fig.~\ref{fig:HHG}b), a distinct peak becomes apparent in the high harmonic spectra at a frequency corresponding to the excitonic energy  $E_{\rm exciton}$, marked by the pink dashed line.  Subsequently, there is a discernible uptick in HHG around this peak, followed by the initiation of a plateau that commences at the charge or Mott gap (indicated by the orange line). This intriguing observation aligns with the findings documented in \cite{udono_excitonic_2022}. The excitonic peak, as noted previously, originates from the $V$-induced binding of the single doublon-holon pair and hence has a lower energy than the Mott gap. Its appearance in the high harmonic spectrum indicates the partial shift of the spectral weight from the free charge carriers to the exciton.

As $V$ exceeds $U/3$, a notable shift occurs, giving rise to the production of multiple doublon-holon pairs bound together to form $n_{\text{ex}}$-excitons. In the specific case of $U/t_0 = 30$ and $V/t_0 = 12$, below the Mott gap, a significant augmentation in the HHG spectra is evident around energies corresponding to excitations comprised of multiple doublon-hole pairs, such as biexcitons (Fig.~\ref{fig:HHG}c). Notably, the excitonic peak is marked by the dashed pink line, occurring at a considerably lower energy. Simultaneously, at a higher yet still sub-Mott gap energy level, a bi-exciton is discernible in the spectra, indicated by the dashed purple line. This exploration underscores that by adjusting the parameter $V$, we have the ability to induce heightened HHG peaks precisely localized around specific $n_{\text{ex}}$-exciton energies. The ability to finely control such excitonic features through parameter tuning opens avenues for tailoring the non-linear HHG response of the system at particular frequencies.

\paragraph{Critical regime.}
In the proximity of the critical point $(U-V)/t_0 < 2$ in the Mott phase, where size fluctuations become increasingly likely, a noteworthy phenomenon emerges — the formation of CDW droplets. This intriguing development is vividly depicted in the HHG spectrum, showcasing a distinctive plateau below the Mott gap (Fig.~\ref{fig:HHG}d). This plateau initiates around a frequency corresponding to the energy $\sim U-V-4t_0$ (indicated by the dashed cyan line), marking the onset of a continuum of CDW droplet states. The non-linear optical response from this continuum, composed of excitonic strings of various sizes, gives rise to the observed plateau below the Mott gap (highlighted by the orange line). This subtle observation helps to differentiate the nature of excitations within the system, emphasizing the importance of understanding the dynamic behavior near the critical point in such a system. Furthermore, the observed enhanced HHG response, as compared to the weak signals in the linear response, yields a significant advantage.
%Charg
\paragraph{CDW phase.}
We now devote our attention to the case where $2V>U$, which means that the ground state of the system has a spontaneously broken translational symmetry of the lattice and exhibits the CDW order. 
An exemplary case that vividly illustrates the complexity introduced by the CDW order is the HHG spectrum of the ``pure'' ($U=0$) CDW phase, shown in Fig.~\ref{fig:HHG}e). The CDW phase brings forth a distinctive optical landscape, considerably more intricate than that of the "pure" ($V=0$) Mott insulator, owing to the various possible optical excitations present in this regime.
Notably, the CDW phase is characterized by the opening of an additional spin gap $\equiv E_0(N_\uparrow+1,N_\downarrow-1)+E_0(N_\uparrow-1,N_\downarrow+1)-2E_0(N_\uparrow,N_\downarrow)$ which is the gap to spin excitations. Intuitively, the CDW ground state at $U/t_0,V/t_0 \gg 1$, consists of doublon-holon pairs arranged alternately on each site.  A spin flip on a doublon site would be impossible due to Pauli exclusion, and therefore, the flipped spin must occupy an empty site leaving behind a singlon in one of the doublon sites. Thus, the presence of neighboring interactions would incur an energy cost $\sim V$ which implies that there is a gap to spin excitations. 

The spin gap thus corresponds to the excitations with one broken doublon-hole pair with two created singlons. When such singlons are bound together, they are referred to as spin density wave (SDW) droplets. Ignoring their delocalization over $L$ sites for low values of $t_0$, their excess energies are given by $\sim V-n(U-2V)$, where $n$ is a droplet size \cite{hirsch_charge_1984}. The lowest energy droplet (for $V$ higher than the critical value $V>U/2$) is a droplet of size $n=1$ and has energy $\sim 3V-U$ (pink dashed line) that corresponds to the spin gap (green solid line). On the other hand, the charge gap (thick orange line) corresponds to the two delocalized free singlons, coming from the one broken pair, and has an energy $\sim 4V-U$ (green ochre dashed line). In the HHG spectrum, there are also small peaks at higher energies -- centered around the frequency corresponding to $5V-2U$  -- coming from the excitations made up of two broken pairs, bound together and forming a magnetic domain. They correspond to SDW droplets of size $n=2$. The signatures of both the gaps can also be observed in a more realistic CDW scenario, where $U>V$ (Coulomb repulsion stronger on the same site than between neighboring sites), see Fig.~\ref{fig:HHG}f). The additional peaks centered around $5V$ cannot be observed in this case, as their energy is higher than the charge gap only when $V>U$. 

Therefore, in essence, HHG not only captures the presence of CDW or SDW droplets on either side of the transition but also traces the system's transition from one phase to the other through nucleation (formation of clusters of the new phase within the existing phase), signifying the occurrence of a first-order transition. Therefore, this comprehensive analysis also reveals the potential of HHG as a powerful tool for probing and understanding the dynamics of phase transitions in the strongly correlated regime.

\subsection{Subcycle analysis}
\label{s:subcycle}

\begin{figure*}[ht]
    \centering
    \includegraphics[width=2\columnwidth]{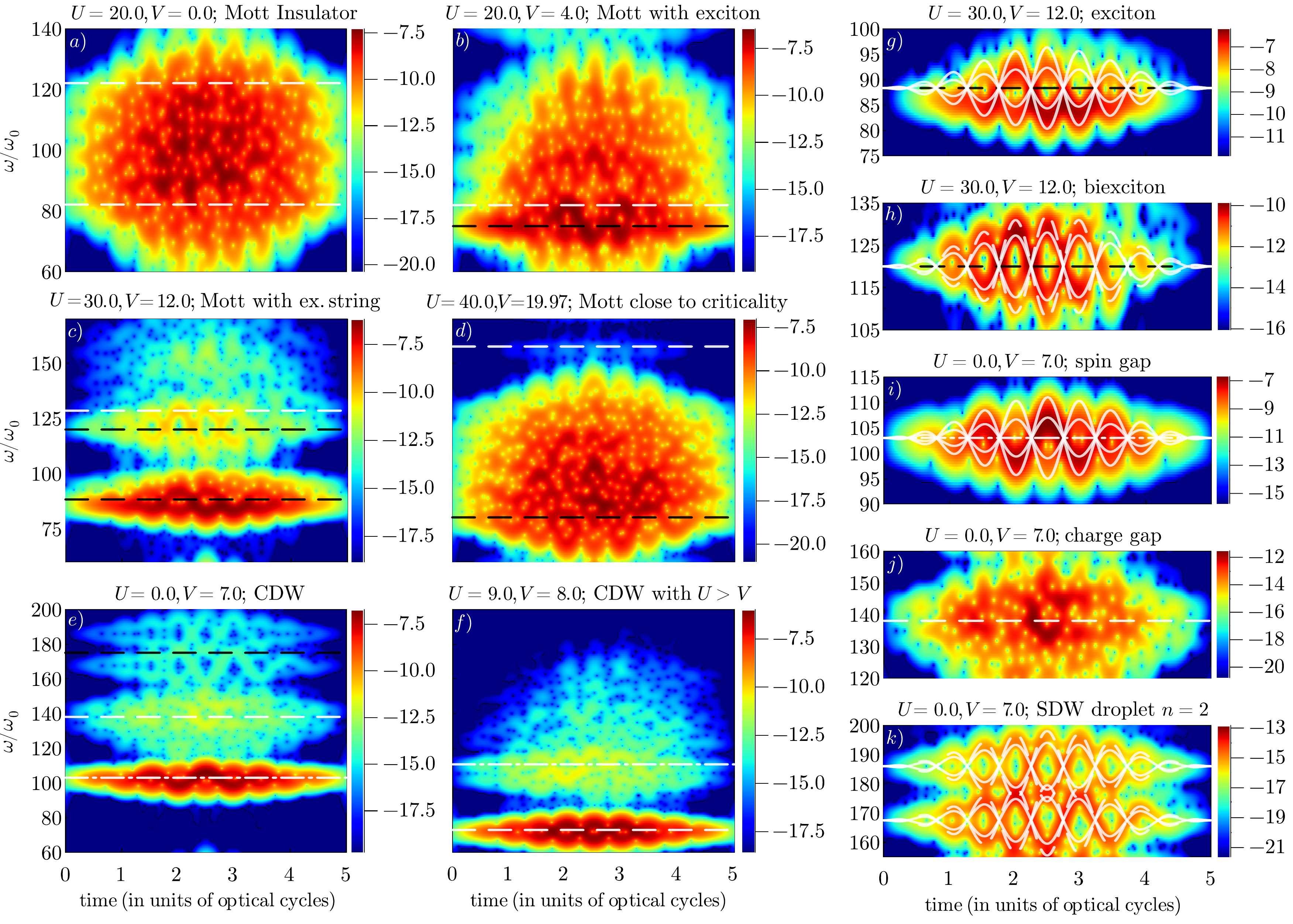}
    \caption{Subcycle analysis of HHG spectra, supplementing the raw HHG data presented in Fig.~\ref{fig:HHG}, the color encodes the value of $\log_{10}(P)$. The left and middle panels show the whole spectra above the optical gap. The dashed white lines are charge gap and, for subplot a) charge gap $+8t_0$, dashed black lines -- exciton, biexciton, $U-V-4t_0$ and SDW droplet of size $n=2$ (compare with Fig.~\ref{fig:HHG}), white dash-dot line -- spin gap. The plots on the right panels show characteristic parts of the full spectra with oscillations proportional to $E(t)$ and $-E(t)$, see the Sec.~\ref{s:subcycle} for details.}
   % COMMENT
   % {\color{red} %[Comment: The panels need letters a),b),..., as referenced in the main text. Shouldn't there be black lines in U=30,V=12 plot left side? Which excitation is the one at $5V \pm 2t_0$? Where are they discussed?
   % Maybe it would be better to arrange the panels in the left and middle columns in the same way as in Figure 2, and then put the panels from the right column as a third line?
   % ]}} \textcolor{green}{Marlena: I think it would be difficult to rearrange the plots this way while keeping it all compact and using all the space, or maybe you have some idea?
   
    \label{fig:Gabor}
\end{figure*}

To delve deeper into the electron dynamics, we turn our attention towards subcycle analysis - a methodology designed to dissect the temporal intricacies of laser-driven electron dynamics at the sub-optical-cycle level. Within the domain of HHG, where electrons undergo ionization and subsequent return to the parent ion, subcycle analysis becomes instrumental in understanding the precise timing and phase relationships governing the emission of high-energy harmonics within a single optical cycle. The results of the subcycle analysis are shown in Fig.~\ref{fig:Gabor}

\paragraph{Free carrier regime.}
In the free charge carrier optical regime, dominated by a freely moving single doublon-holon pair, we observe that the plateau above the Mott gap (marked by the lowest dashed white line) consists of many interfering trajectories. This makes it difficult to easily track individual oscillations, Fig.~\ref{fig:Gabor}a). In the Methods sections, we show to gain insight to this regime by restricting the dynamics of the system. 

\paragraph{Excitonic regime.}
On the contrary, the subcycle analysis in proximity to excitonic and biexcitonic energies reveals distinct oscillations around the respective energy levels, as illustrated in Fig.~\ref{fig:Gabor}b) and c). Two trajectories closely align with the shape of the electric field $E(t) = E_0f(t) = -\partial_t A(t)$ and $-E(t)$. The amplitude of excitonic oscillations lies within the range of $E_0$ to $2E_0$ (solid lines), while for the biexciton with higher dipole moments, it is approximately twice as high, spanning between $2E(t)$ (solid line) and $3E(t)$ (dashed line), as depicted in Fig.~\ref{fig:Gabor}g) and h). 

\paragraph{CDW droplets.}
In the regime of CDW droplets, there is a noteworthy phenomenon of interference among multiple trajectories at frequencies corresponding to the plateau formed below the Mott gap, as depicted in Fig.~\ref{fig:Gabor}d). It appears that optical excitations originating from a continuum of states, such as free charge carriers and CDW droplets, amalgamate to form a plateau featuring multiple interfering trajectories in the nonlinear response. In contrast, excitations that manifest as peaks in the linear response, such as excitons and excitonic strings, are distinctly observed in the nonlinear response as clear oscillations proportionate to the electric field.

\paragraph{CDW phase.} The subcycle analysis of the CDW phase unveils various time-dependent responses, each intricately associated with the distinctive peaks/plateaus observed in the HHG spectrum, as elucidated in Fig.~\ref{fig:Gabor}e) and f). The oscillations linked to the spin gap exhibit a compelling fluctuation of amplitude, ranging between $E_0$ and $2E_0$ (Fig.~\ref{fig:Gabor}i). Simultaneously, excitations associated with the charge gap contribute to the creation of a complex interference-plateau structure, as visualized in Fig.~\ref{fig:Gabor}j). This intricate temporal behavior highlights the intricate interplay of charge and spin dynamics within the CDW phase. Furthermore, the subcycle analysis unveils peaks around $5V$ that manifest themselves as oscillations with amplitudes ranging from $2E_0$ to $3E_0$ (Fig.~\ref{fig:Gabor}k). This distinctive pattern sheds light on the diverse nature of excitations occurring within the CDW phase, offering a detailed perspective on the underlying processes that contribute to the observed harmonic generation.

%This approach aids in 

\section{Discussion}

Our results have elucidated the quantum processes governing the interaction of intense laser fields with matter. We have shown that HHG spectroscopy can not only locate the presence of the various optical many body excitations, but through subcycle analysis it also enables us to probe the structure of the excitations based on the response of the trajectories of their constituents to the incident electric field of the laser. From this perspective, it has become evident that the subcycle analysis serves as a powerful tool for unraveling the intricate trajectories responsible for the formation of peaks or plateaus, characterizing diverse excitations across the optical spectra in the four distinct regimes. When focusing on isolated peaks, the analysis discerns a noteworthy correlation: excitations characterized by higher charges or increased dipole moments manifest with proportionally elevated amplitudes of oscillation. This insightful observation allows for decoding the nature and complexity of the excitations, distinguishing whether the observed peaks originate from single excitons or extend to higher-order states such as biexcitons, triexcitons, or even more complex configurations. Remarkably, this approach extends its utility to the realm of spin droplets as well, particularly in the parameter space associated with the CDW phase.

%{\color{red} [I think we should keep this much shorter, and/or we may consider presenting a figure with linear optics spectrum? Just a long text is not so nice.]}
% {\color{blue} [ 
 To conclude let us emphasize that insight obtained from non-linear optics goes far beyond the information revealed by the linear optical conductivity.
Linear optical response,  excels at identifying simple excitations like free charge carriers (holon-doublon pairs) and tightly bound excitons. In the context of strongly correlated systems often represented by the EFHM studied here, this translates to a continuous band in the optical spectrum marking the Mott gap and a prominent delta peak at a specific energy corresponding to the tightly bound exciton state. These features arise because the linear response regime deals with weak perturbations to the ground state. The creation of such excitations requires minimal energy investment, and their well-defined electronic states lead to strong optical transitions easily captured by linear probes.
However, the limitations of linear response become apparent when investigating more complex phenomena like excitonic strings and CDW droplets. These features involve multiple excitations interacting and forming extended structures. In the case of excitonic strings, the creation of each additional exciton within the string requires overcoming a smaller energy barrier compared to the initial exciton formation. While this reduced energy cost facilitates string formation, the transition matrix elements between the ground state and higher-order excited states containing excitonic strings become significantly weaker. These weak matrix elements translate to faint peaks in the optical conductivity spectrum, making them challenging to detect with linear response methods.

This is where non-linear optical response becomes advantageous. By employing stronger external fields, non-linear techniques can access higher-order excitations. When dealing with excitonic strings, non-linear approaches can overcome the weak ground state to excited state transitions observed in linear response. Additionally, they can explore the interplay between multiple excitons within the string, providing a deeper understanding of their binding energies and spatial configurations.
Similarly, non-linear methods can shed light on CDW droplets as well and can potentially reveal the characteristic energy scales associated with CDW droplet formation and their size dependence.

In summary, we have shown that the high-harmonic generation can serve as a tool for probing optical excitations in strongly correlated systems. It has been demonstrated that the strongly nonlinear response of the extended Fermi Hubbard model contains information about different excitonic patterns, both in the Mott and CDW phases. Moreover, the observed changes in the high harmonic spectrum, resulting from the increased nearest-neighbor interaction strength, make it possible to track the formation of clusters, a process that governs the first-order phase transition. Subcycle analysis provides additional insight into the properties of excitons at given energies, allowing one to distinguish between exciton chains or SDW droplets of different sizes and to decode whether the contribution originates from a continuum of states or from single peaks in the linear response of the system.

\section{Methods}
\subsection{Non-linear optical response}
We begin with the system's ground state as our initial condition, obtained through exact diagonalization calculations. Next, we time-evolve this state using the Runge-Kutta method while subjecting it to a laser field described by a vector potential exhibiting a sine-square envelope,
\begin{equation}
A(t) = A_0 \sin^2\left(\frac{\omega_0 t}{2 n_c}\right) \sin\left(\omega_0 t\right),\qquad 0<t<\frac{2\pi n_c}{\omega_0}.
\end{equation}
where $\omega_0$ is a laser frequency, $A_0$ is an amplitude and $n_c$ is the number of cycles. In contrast to some previous works \cite{silva_high-harmonic_2018}, here, we choose a low enough laser amplitude to avoid the light-induced phase transition and the breakdown of an insulating state. We use the velocity gauge and simulate the light-matter coupling via Peierls substitution,
\begin{equation}
    t_0 \hat{c}^{\dagger}_{j\sigma} \hat{c}_{j+1\sigma} \to t_0 e^{iaA(t)} \hat{c}^{\dagger}_{j\sigma} \hat{c}_{j+1\sigma}.
\end{equation}
The intensity of the HHG spectrum $P(\omega)$ is given by a Fourier transform of the time derivative of the current operator proportional to the acceleration of charges in the system,
\begin{equation} \label{eq:power}
P(\omega) = \left| \mathrm{FT_W}\left[\braket{\dot{\hat{J}}(t)}\right]\right|^2,
\end{equation}
where
\begin{equation}
    \hat{J}(t) = iat_0\sum_{j,\sigma} \left[ e^{iaA(t)} \hat{c}^{\dagger}_{j\sigma} \hat{c}_{j+1\sigma} - e^{-iaA(t)} \hat{c}^{\dagger}_{j+1\sigma} \hat{c}_{j\sigma}  \right].
\end{equation}

\subsection{Subcyle analysis}
The results of the subcycle analysis are obtained through the Gabor transform,
\begin{equation}
   P(\omega, \tau) = \left| \int_{0}^{2\pi n_c/\omega_0} e^{-(t-\tau)^2/\sigma^2} e^{-i\omega t} \braket{\dot{\hat{J}}(t)} dt \right|^2
\end{equation}
with a sliding window width $\sigma = 1/(2\omega_0)$, see Fig.~\ref{fig:Gabor}. %{\color{red} [Comment: Change $\tau$ by $t$ to agree with the notation in Fig.~5?]}

\subsection{Linear response}
The linear optical response of the system is captured by the optical conductivity,
\begin{equation}
    \sigma (\omega>0) = \frac{1}{\omega}\mathrm{Im} \left\{\frac{1}{aL}\sum_n \frac{-\left| \braket{0|\hat{J}_0|n} \right|}{\hbar\omega -(E_n-E_0)+i\eta}\right\}.
\end{equation}
Here, $a$ is the lattice constant, $L$ is the size of the system, and the sum goes over the eigenstates $\ket{n}$ of Hamiltonian \eqref{EFHM}, corresponding to the energies $E_n$. The contribution $i\eta$ broadens the signal, and the current operator $\hat{J}_0$ is given by 
\begin{equation}
    \hat{J}_0 = iat_0\sum_{j,\sigma} \left( \hat{c}^{\dagger}_{j\sigma} \hat{c}_{j+1\sigma} - \hat{c}^{\dagger}_{j+1\sigma} \hat{c}_{j\sigma}  \right).
\end{equation}

\subsection{Finite size effects}
\begin{figure}[!h]
    \centering
    \includegraphics[width=1\columnwidth]{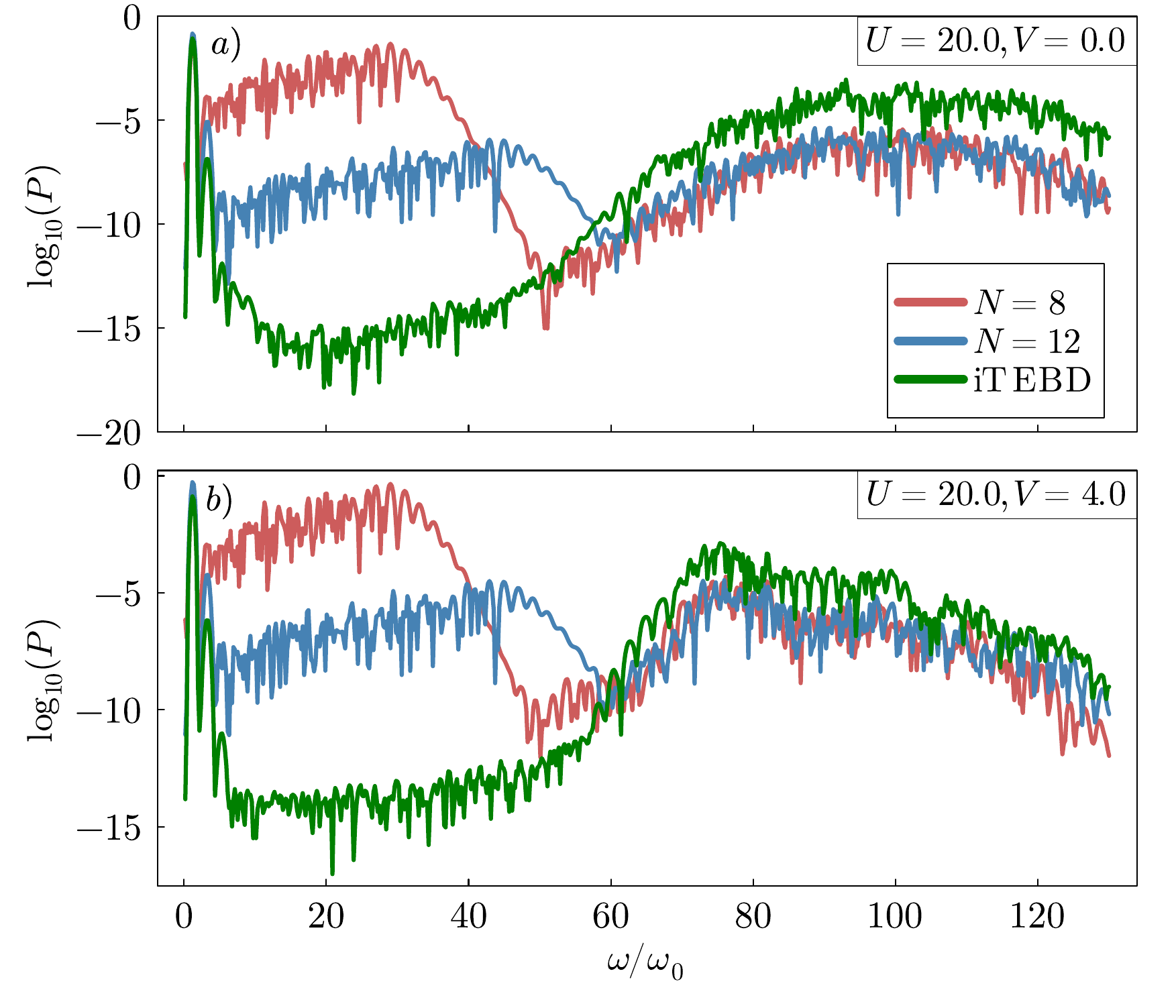}
    \caption{Finite size analysis for two sets of parameters of the Hamiltonian \eqref{EFHM}, laser parametrs: $A_0=4$, $\omega_0=0.2$, $n_c = 5$.}
    \label{fig:fs_Mott}
\end{figure}

\begin{figure}[!h]
    \centering
\includegraphics[width=1\columnwidth]{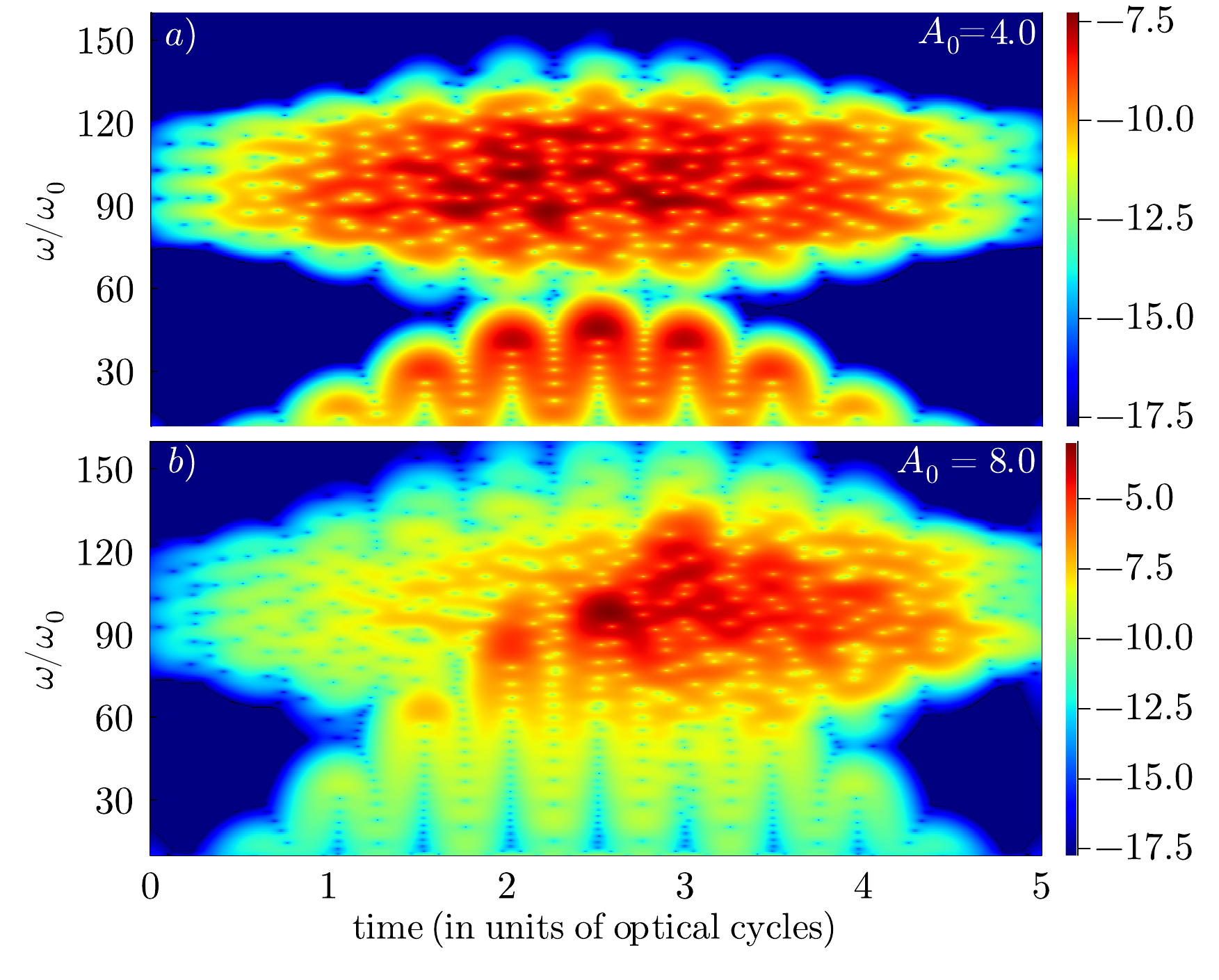}
    
    \caption{Finite size effect in Gabor transform for two laser field intensities: a) $A_0 = 4.0$ and b) $A_0 = 8.0$ for the Mott insulator with parameters $U=20.0$ and $V=0.0$. The color encodes the value of $\log_{10}(P)$}
    \label{fig:fs_Gabor}
\end{figure}

Here, we compare the results of exact calculations for two system sizes $N=8$ and $N=12$ as well as results obtained in the time-evolution block-decimation method with infinite boundary conditions (iTEBD), see Fig.~\ref{fig:fs_Mott} for examples. We observe that the finite system affects the low-frequency nonlinear response of our model by introducing an additional peak below the optical gap of the system. The intensity of this peak decreases with system size, and finally it disappears in the thermodynamic limit represented by the iTEBD calculations. The subcycle analysis shows a clear oscillating trajectory proportional to the $|E(t)|$, see Fig.~\ref{fig:fs_Gabor} In the main part of the paper, we decided to include exact finite-size results due to their precision even in high frequency ranges and long evolution times, where convergence of iTEBD calculations is computationally very demanding. 

\subsection{Separating dynamical processes in HHG}

\paragraph{Restricted subcycle analysis of unbound doublon-hole pair.}
 The kinetic part of the Hamiltonian \eqref{EFHM}, as well as the current operator, can be rewritten using
\begin{align}
    \hat{c}^{\dagger}_{j\sigma} \hat{c}_{k\sigma} &=\hat{c}^{\dagger}_{j\sigma} (\mathbb{I} - \hat{n}_{j\sigma'}+\hat{n}_{j\sigma'}) \hat{c}_{k\sigma}(\mathbb{I} - \hat{n}_{k\sigma'}+\hat{n}_{k\sigma'}) \nonumber \\
    &=[\hat{c}^{\dagger}_{j\sigma} \hat{h}_{j\sigma'} +\hat{c}^{\dagger}_{j\sigma}\hat{n}_{j\sigma'} ][\hat{c}_{k\sigma}\hat{h}_{k\sigma'} +\hat{c}_{k\sigma}\hat{n}_{k\sigma'} ] \nonumber \\
    &=\hat{c}^{\dagger}_{j\sigma} \hat{h}_{j\sigma'} \hat{c}_{k\sigma} \hat{n}_{k\sigma'} +\hat{c}^{\dagger}_{j\sigma} \hat{n}_{j\sigma'} \hat{c}_{k\sigma} \hat{h}_{k\sigma'}
    \label{pair_cr_ann} \\
    &+\hat{c}^{\dagger}_{j\sigma} \hat{h}_{j\sigma'} \hat{c}_{k\sigma} \hat{h}_{k\sigma'} +\hat{c}^{\dagger}_{j\sigma} \hat{n}_{j\sigma'} \hat{c}_{k\sigma} \hat{n}_{k\sigma'},
    \label{d_h_hopping}
\end{align}
where $\hat{h}_{j\sigma}= \mathbb{I} - \hat{n}_{j\sigma}$.
The first two terms of the sum above \eqref{pair_cr_ann} correspond to the annihilation and creation of a doublon-holon pair, while the terms in \eqref{d_h_hopping} correspond to the hopping of a holon and a doublon, respectively. This dynamics separation bears a resemblance to the 3-step model in Mott insulator described in \cite{murakami_high-harmonic_2021}.
%\textcolor{blue}{[Do you think it would be more useful to have a discussion in terms of intraband and interband processes, where the last two terms describe the lower and upper band hoppings respectively?]. The   terms in Eqn.(\ref{d_h_hopping}) are doublon conserving terms and correspond to hopping in the lower and upper Hubbard bands, while \eqref{pair_cr_ann}  consists of intra-Hubbard-band terms.} 
%\textcolor{blue}{[Comment: Perhaps we can describe a "3 step" model in terms of the Hubbard bands? For a large U Mott phase like here, this would be especially appealing and natural.]}

\begin{figure}[h]
    \centering
    \includegraphics[width=1\columnwidth]{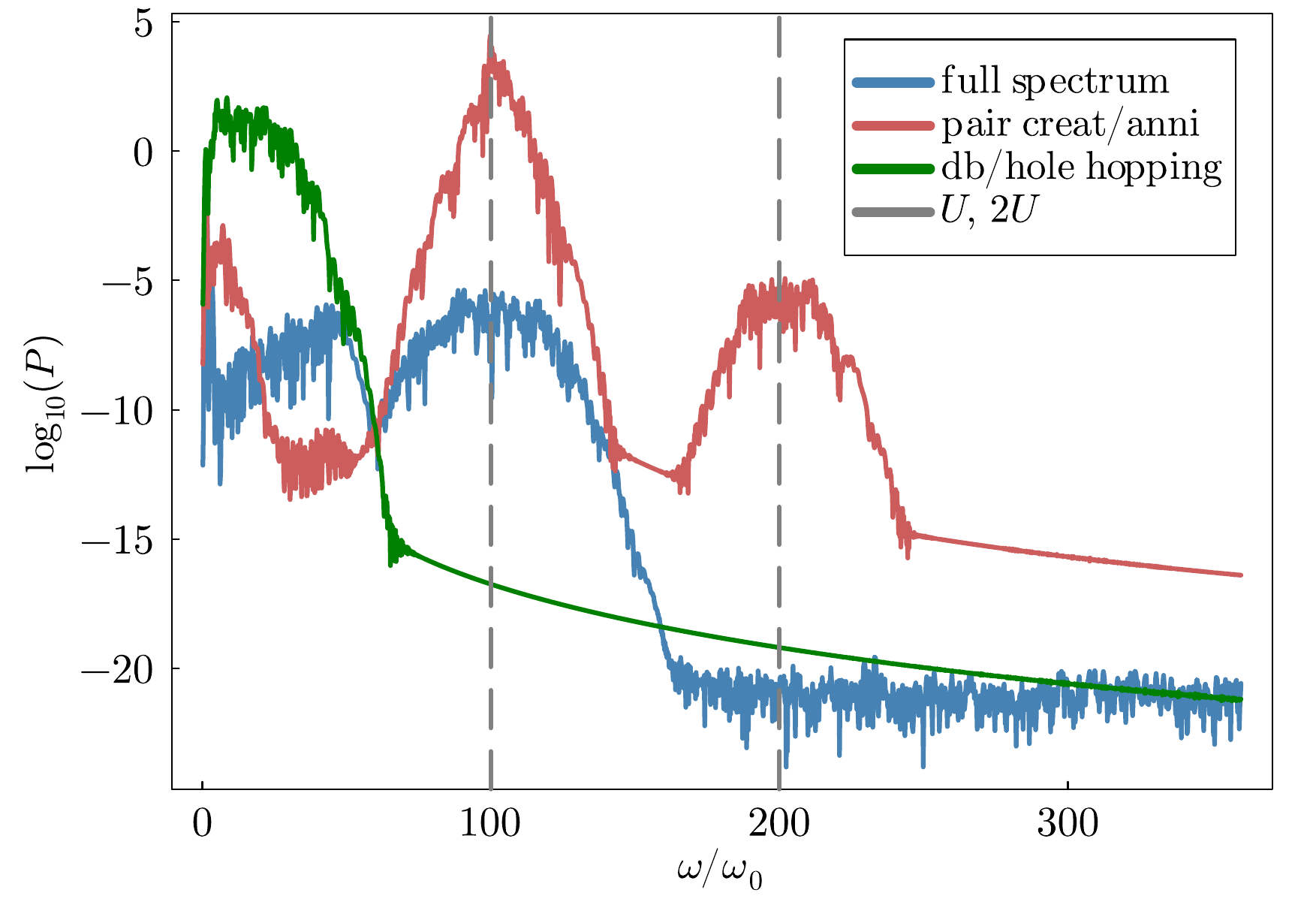}
    \caption{High-harmonic spectra of the full Hamiltonian and the truncated Hamiltonians [according to Eq.~(\ref{pair_cr_ann}) or (\ref{d_h_hopping}] at   ${V=0.0}, \; {U=20.0}, \; A_0=4.0, \; \omega_0 = 0.2, \; n_c = 5$. Vertical lines correspond to $U$ and $2U$
    }
    \label{fig:HHG_sep}
\end{figure}

\begin{figure}[h]
    \centering
    \includegraphics[width=1\columnwidth]{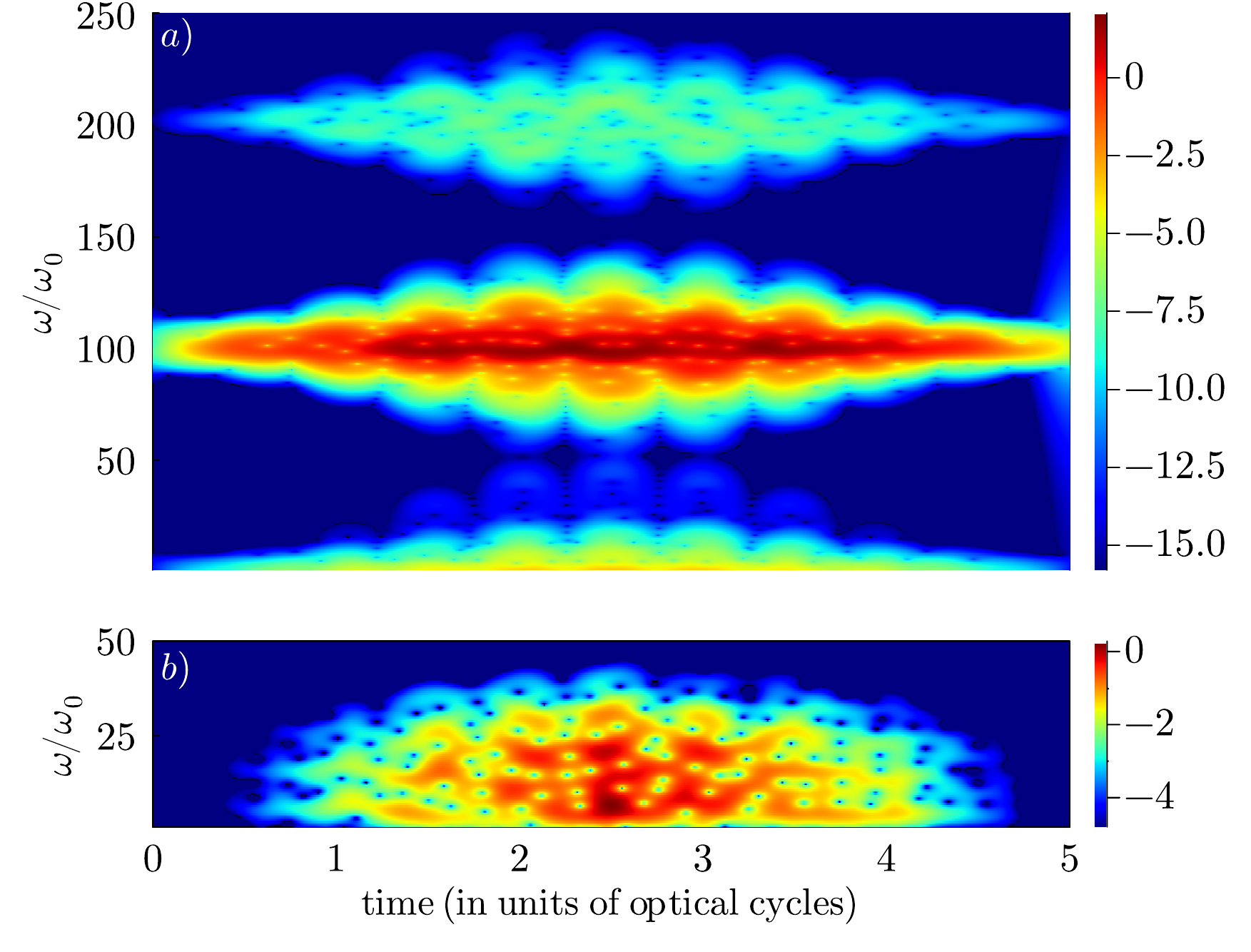}
    \caption{Subcycle analysis of the HHG spectra presented in the Fig.\ref{fig:HHG_sep}, the color encodes the value of $\log_{10}(P)$. The a) panel corresponds to the doublon-hole pair creation and annihilation processes, as in Eq.\eqref{pair_cr_ann}, and the b) panel corresponds to the hopping of a doublon or hole, as in Eq.\eqref{d_h_hopping}.
    %{\color{red} [Comment: This caption needs to be expanded. The color scheme depicts $P(\omega)$, right? Units for $P$ and time are missing.]}
    }
    \label{fig:Gabor_sep}
\end{figure}

We performed calculations on the high harmonic spectra of the Mott phase, constraining electronic hoppings solely to the terms specified in \eqref{d_h_hopping} or \eqref{pair_cr_ann}.
%{\color{red} [Comment: Do I understand correctly: either (9) or (10)?]}\textcolor{blue}{[So as said in the previous paragraph,   either (9) or (10) are considered in both H and J, right?]}
 Refer to Fig.~\ref{fig:HHG_sep} for the HHG spectra and Fig.~\ref{fig:Gabor_sep} for the subcycle analysis. It is important to note that the initial state of the evolution remains the ground state of the full Hamiltonian. 
%\textcolor{blue}{[Would it not be useful to keep the full H during the evolution, and distringuish between the different correlated band processes by using (9) and (10) only in J? ]} 

When dynamics are confined to doublon and holon hopping terms, a single below-gap plateau is observed, as per expectations. In the subcycle analysis, this plateau manifests itself as multiple interfering trajectories. Conversely, when only pair creation/annihilation processes are allowed and the resulting HHG spectrum solely arises from the recombination of doublon-hole pairs, three distinct peaks emerge. The first, below-gap peak is attributed to effective hopping resulting from the simultaneous annihilation and creation of a pair at different lattice sites. The highest and dominant peak, centered around $U$, can be explained as the recombination of a single doublon-hole pair. The third peak, centered around $2U$ probably originates from the recombination of two pairs simultaneously, a phenomenon suppressed in the full spectrum derived from non-restricted dynamics. Interestingly, the overall intensity of the spectrum increases for the restricted dynamics of the system. This analysis provides insight on the nature and origin of plateaus seen in the optical regime with free charge carriers.

\paragraph{Interband vs. intraband contributions}
 In the case of band insulators, the current can be split into the intra- and inter-band contributions in the following way,
\begin{align}
\label{eq:intra_inter}
   J_{\textrm{tot}} &= J_{\textrm{inter}} + J_{\textrm{intra}},\nonumber \\
   J_{\textrm{inter}} &= \sum_{b}\sum_{n,m} \langle \psi(t) | \phi^{b}_n \rangle\langle \phi^{b}_n | \hat{J} | \phi^{b}_m \rangle\langle \phi^{b}_m | \psi(t)\rangle, \nonumber \\
   J_{\textrm{inter}} &= \sum_{b,p\neq b}\sum_{n,m} \langle \psi(t) | \phi^{b}_n \rangle\langle \phi^{b}_n | \hat{J} | \phi^{p}_m \rangle\langle \phi^{p}_m | \psi(t)\rangle,
\end{align}
where $|\phi_n^b\rangle$ is a $n$-th eigenstate from band $b$.
The intraband current includes the transitions between states from the same band, whereas the interband current involves transitions between different bands. In the case of single-active-electron picture, it was shown that the intraband contribution is dominant for lower-order harmonics and the interband current dominates in the higher-harmonic region \cite{guan_high_2016}.
\\
However, we realize that deciphering the processes that govern both the low- and high-order harmonic regimes in strongly correlated systems poses a greater challenge due to the absence of a well-defined band structure. Nevertheless, by analogy, a three-step model for Mott insulators akin to semiconductors can still be proposed. In this model, the creation of doublon-hole pairs is analogized to electron tunneling to the conduction band, doublon-hole hopping mirrors intraband current, and doublon-hole annihilation represents the recombination of electrons and holes, consequently giving rise to interband harmonics. 
\\
In this scenario, the formulas in Eq.~\eqref{eq:intra_inter} are not applicable to disentangle these processes. Instead, an alternative approach involves projecting the current operator onto the subspace in the position basis, rather than the eigenstate basis. This allows for the identification of a subspace comprising basis states with a specific number of doublon-hole pairs:
\begin{align}
    J_{\textrm{hopping}} &= \sum_{i} \langle \psi(t)| \hat{P}_i \hat{J} \hat{P}_i |\psi(t)\rangle, \nonumber \\
    J_{\textrm{cr/an}} &= \sum_{i,j\neq i} \langle \psi(t)| \hat{P}_i \hat{J} \hat{P}_j |\psi(t)\rangle,
\end{align}
where $\hat{P}_i$ represents a projector onto a subspace with $i=0, 1, ..., N$ doublon-holon pairs. It is important to note that this approach does not impose constraints on the evolution of the initial state, which may not conserve the number of doublon-holon pairs.

\begin{acknowledgments}
T.G. acknowledges funding by the Agencia Estatal de Investigación (AEI) through Proyectos de Generación de Conocimiento PID2022-142308NA-I00 (EXQUSMI), and that the work has been produced with the support of a 2023 Leonardo Grant for Researchers in Physics, BBVA Foundation. The BBVA Foundation is not responsible for the opinions, comments and contents included in the project and/or the results derived therefrom, which are the total and absolute responsibility of the authors. U.B. acknowledges the project that gave rise to these results, received the support of a fellowship (funded from the European Union’s Horizon 2020 research and innovation programme under the Marie Sklodowska-Curie grant agreement No 847648) from “la Caixa” Foundation (ID 100010434). The fellowship code is “LCF/BQ/PR23/11980043”.
RWC acknowledges support from the Polish National Science Centre (NCN) under the Maestro Grant No. DEC-2019/34/A/ST2/00081. M.M.M. acknowledges that this research is part of the project No. 2021/43/P/ST3/03293 co-funded by the National Science Centre and the European Union’s Horizon 2020 research and innovation programme under the Marie Sk\l{}odowska-Curie grant agreement no. 945339. M.D. acknowledges support from the National Science Centre (Poland) under Grant No. 2022/04/Y/ST3/00061. M.L.B. acknowledges financial support from the Spanish MCIN/AEI/10.13039/501100011033 grant PID2020-113334GB-I00, Generalitat Valenciana grant CIPROM/2022/66, the Ministry of Economic Affairs and Digital Transformation of the Spanish Government through the QUANTUM ENIA project call - QUANTUM SPAIN project, and by the European Union through the Recovery, Transformation and Resilience Plan - NextGenerationEU within the framework of the Digital Spain 2026 Agenda, and by the CSIC Interdisciplinary Thematic Platform (PTI+) on Quantum Technologies (PTI-QTEP+). This project has also received funding from the European Union’s Horizon 2020 research and innovation program under grant agreement CaLIGOLA MSCA-2021-SE-01-101086123.

ICFO group acknowledges support from: ERC AdG NOQIA; MICIN/AEI (PGC2018-0910.13039/501100011033, CEX2019-000910-S/10.13039/501100011033, Plan National FIDEUA PID2019-106901GB-I00, FPI; MICIIN with funding from European Union NextGenerationEU (PRTR-C17.I1): QUANTERA MAQS PCI2019-111828-2); MCIN/AEI/10.13039/501100011033 and by the ``European Union NextGeneration EU/PRTR'' QUANTERA DYNAMITE PCI2022-132919 (QuantERA II Programme co-funded by European Union’s Horizon 2020 programme under Grant Agreement No 101017733), Proyectos de I+D+I ``Retos Colaboración''
QUSPIN RTC2019-007196-7); Fundació Cellex; Fundació Mir-Puig; Generalitat de Catalunya (European Social Fund FEDER and CERCA program, AGAUR Grant No. 2021 SGR 01452, QuantumCAT U16-011424, co-funded by ERDF Operational Program
of Catalonia 2014-2020); Barcelona Supercomputing Center MareNostrum (FI-2023-1-0013); EU Quantum Flagship (PASQuanS2.1, 101113690); EU Horizon 2020 FET-OPEN OPTOlogic (Grant No 899794); EU Horizon Europe Program (Grant Agreement 101080086 — NeQST), National Science Centre, Poland (Symfonia Grant No. 2016/20/W/ST4/00314); ICFO Internal ``QuantumGaudi'' project; European Union’s Horizon 2020 research and innovation program under the
Marie-Skłodowska-Curie grant agreement No 101029393 (STREDCH) and No 847648 (“La Caixa” Junior Leaders
fellowships ID100010434: LCF/BQ/PI19/11690013, LCF/BQ/PI20/11760031, LCF/BQ/PR20/11770012, LCF/BQ/PR21/11840013).  Views and opinions expressed are, however, those of the authors only and do not necessarily reflect those of the European Union, European Commission, European Climate, Infrastructure and Environment Executive Agency (CINEA), nor any other granting authority. Neither the European Union nor any granting authority can be held responsible for them.
U.B. is also grateful for the financial support of the IBM Quantum Researcher Program.
\end{acknowledgments}

%\clearpage
\bibliography{HHGEFHM.bib}	
\end{document}